\documentstyle[prc,aps,amssymb,epsfig]{revtex}

\begin{document}
\topmargin -20pt

\hbadness=10000

\twocolumn[\hsize\textwidth\columnwidth\hsize
           \csname  @twocolumnfalse\endcsname

\title{n-p Short-Range Correlations from (p,2p + n) Measurements}

\author{A. Tang$^a$, J. W. Watson$^a$,
J. Aclander$^b$, J. Alster$^b$, G. Asryan$^{d,c}$, 
Y. Averichev$^{h}$, D. Barton$^d$, V. Baturin$^{f,e}$, \\
N. Bukhtoyarova$^{d,e}$, 
A. Carroll$^d$, S. Heppelmann$^f$, A. Leksanov$^f$, 
Y. Makdisi$^d$, A. Malki$^b$, E. Minina$^f$, I. Navon$^b$, \\		
H. Nicholson$^g$,  
A. Ogawa$^f$, Yu. Panebratsev$^{h}$, E. Piasetzky$^b$, 
A. Schetkovsky$^{f,e}$, 
S. Shimanskiy$^{h}$, D. Zhalov$^f$}

\address{$^a$Dept. of Physics, Kent State Univ., Kent, OH 44242, U.S.A.\\
$^b$School of Physics and Astronomy, Sackler Faculty of Exact 
  Sciences, Tel Aviv University, Ramat Aviv 69978, Israel \\
$^c$Yerevan Physics Institute, Yerevan 375036, Armenia\\
$^d$Collider-Accelerator Department, Brookhaven National Laboratory, Upton, NY 11973, USA\\
$^e$Petersburg Nuclear Physics Institute, Gatchina, St. Petersburg 188350, Russia \\
$^f$Physics Department, Pennsylvania State University, University Park,
 PA 16801, U.S.A.\\
$^g$Dept. of Physics, Mount Holyoke College, South Hadley, MA 01075, U.S.A.\\
$^h$J.I.N.R., Dubna, Moscow 141980, Russia}

\maketitle

\begin{abstract}

We studied the $^{12}$C(p,2p+n) reaction at beam momenta of 5.9, 8.0 
and 9.0 GeV/c. For quasielastic 
(p,2p) events we reconstructed {\bf p$_f$} the momentum of 
the knocked-out proton before the reaction; {\bf p$_f$} was 
then compared (event-by-event) with {\bf p$_n$}, the measured, 
coincident neutron momentum. For $\vert$p$_n$$\vert$ $>$ k$_F$ = 
0.220 GeV/c (the Fermi momentum) a strong 
back-to-back directional correlation between {\bf p$_f$} and 
{\bf p$_n$} was observed, indicative of short-range n-p correlations. 
From {\bf p$_n$} and 
{\bf p$_f$} we constructed the distributions of 
c.m. and relative motion in the longitudinal direction 
for correlated pairs. After correcting for 
detection efficiency, flux attenuation and solid angle, 
we determined that 49 $\pm$ 13 $\%$ of events with 
$\vert$p$_f$$\vert$ $>$ k$_F$ had directionally correlated neutrons 
with $\vert$p$_n$$\vert$ $>$ k$_F$.
Thus short-range 2N correlations are a major source of high-momentum 
nucleons in nuclei.

\end{abstract}

]     
\begin{narrowtext}

For the past half century, the dominant model for the structure of
nuclei,
especially light nuclei, has been the nuclear shell model. In the shell  
model, the long-range (${\sim}$ 2 fm) part of the N-N force, in   
combination with the Pauli principle, produces an average potential in
which the nucleons undergo nearly independent motion, and the residual
interactions can be treated by perturbation theory. However the NN
interaction is also highly repulsive at short-range (${\sim}$ 0.4 fm).
This short-range two-nucleon repulsion is responsible for the saturation  
density of nuclei, e.g. the nearly constant density in the interior of all
stable nuclei heavier than $^4$He. The effect of this two-nucleon
short-range (2N-SRC) repulsion should also manifest itself in the motions
of the nucleons in the nucleus. This topic has been
actively pursued in recent years with electro-magnetic interactions.
%
Recently, Aclander et al. \cite{acl99} described a new technique
 for observing 2N-SRC with a high momentum transfer proton
reaction, based on a proposal by Frankfurt and Strikman
\cite{LFran1}.
 In this letter we show that this
method not only demonstrates the presence of 2N-SRC but also provides
quantitative information on the strength of 2N-SRC. We also obtained
the first measured results of the c.m. momentum of the
correlated
NN pair as well as the relative momentum of the two nucleons in their
c.m. frame.

 For the quasi-elastic knockout of protons
from nuclei, e.g. $^{12}$C(p,2p) in \cite{YMar} and for this work, we can
use the Plane-Wave impulse approximation (PWIA) to reconstruct
(event-by-event) the three momentum {\bf p$_f$} that the struck target
proton had before the reaction: \begin{equation}
   {\bf p_{\em f}} = {\bf p_{1}} + {\bf p_{2}} - {\bf p_{0}}
\label{pf}
\end{equation}
where {\bf p$_0$} is the momentum of the incident proton and
{\bf p$_1$} and {\bf p$_2$} are the momenta of the two detected
protons. 

For p-p elastic scattering near 90$^\circ$, the cross section ${d\sigma
\over{dt}}$ falls as {\em s$^{-10}$} where {\em s} and {\em t} are the
standard Mandelstam variables for the square of the total c.m. energy and
the squared momentum transfer, respectively. 
From this strong dependence on {\em s} (viz. {\em s$^{-10}$}) we expect p-p
quasi-elastic scattering to occur preferentially with nuclear protons that
have their momenta {\bf p$_f$} roughly parallel to {\bf p$_0$} the beam
momentum, because this reduces {\em s} and increases ${d\sigma \over{dt}}$
\cite{cr1}.  When two nucleons in a nucleus interact at short range, they
must have large, nearly equal, momenta in opposite directions(because of
their strong repulsion at short range) \cite{AShalit}. If the pair is
broken by a quasi-elastic interaction one of the nucleons may emerge from
the nucleus, with relatively large momentum,
 in a direction opposite to original momentum of its partner.
 Our experiment then consists of
measuring the two protons
of the quasi-elastic scattering in triple coincidence with the emerging 
correlated neutron. Since {\bf p$_f$} is roughly parallel to the beam, we
 placed 36 neutron
detectors primarily in the backward hemisphere, to search for neutrons
with {\bf p$_n$} $\approx$ $-${\bf p$_f$}.  The experimental details were
described in \cite{Malki}.

The measurement reported here (and in \cite{acl99}) were taken with the  
EVA spectrometer \cite{Malki,wu} at the AGS accelerator at Brookhaven
National Laboratory. EVA consists of a 0.8 T superconducting solenoid,  
3.3 m long and 2 m in diameter, with the beam incident along the central
axis. Coincident pairs of high transverse-momentum ({\bf p$_t$}) protons
are detected with four concentric cylinders of strawtube drift chambers.  
EVA was designed for studying quasi-elastic (p,2p) reactions near
90$^\circ$ c.m. and to investigate nuclear transparency \cite{IMardor,ct}.

For all of the results discussed below we applied the four following cuts 
to the data:
\newcounter{bean}
\begin{list}%
{\ \ \ \ \ \ \ (\arabic{bean}).}{\usecounter{bean} \setlength{\rightmargin}{\leftmargin}}
 \item  The projectile should be a proton, as determined by 
\v{C}erenkov detectors in the beam.
 \item  There should be two (and only two) high-{\bf p$_t$} ($>$ 0.6 GeV/c) 
positive-charge tracks.
 \item  The missing energy, E$_{miss}$ should be appropriate for 
quasi-elastic scattering, and was defined by the resolution at 
each energy \cite{YMar}.
 \item  The neutron momentum should be in the range: 0.05 $<$ p$_n$ 
$<$ 0.55 GeV/c.
 \item  The laboratory azimuthal angles for the two detected protons must 
lie within $\pm$ $45^\circ$ of the plane through EVA parallel to the face 
of the active neutron detector. 
\end{list}

Cut (5) was implemented so that we could combine our results with those 
in \cite{acl99}, which were also 
taken with EVA but with more limited acceptance.

\begin{figure}[b]  
\vspace{-.6cm}
\centerline{\epsfxsize=0.34\textwidth \epsfbox{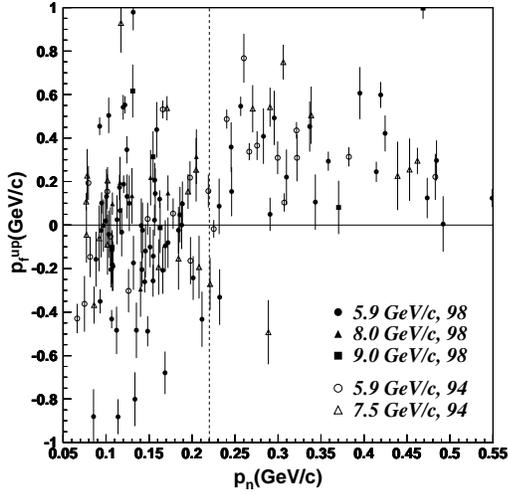}}
\vspace{-.6cm}
\caption{ Scatter plot of p$_f^{up}$ vs. p$_n$ with cuts 1, 2, 3, 4 and 5 
(see text) for $^{12}$C(p,2p+n) events. Data labelled ``98'' (solid symbols) 
are 
for this work. Data labelled ``94'' are from 
Aclander, et al. (Ref.\protect\cite{acl99}). 
The vertical dashed line at 0.22 GeV/c corresponds to k$_F$, 
the Fermi momentum for $^{12}$C. }
\vspace{.2cm}
\label{PfxPn45h}
\end{figure}

Our primary objective was to identify short-range correlated n-p 
pairs in $^{12}$C nuclei and to determine their properties. 
The signature of such pairs is that the 
momentum of the detected neutron, {\bf p$_n$} $\approx -${\bf p$_f$}.
We should expect, however, that the motion of the c.m. of such an n-p 
pair in the nucleus will keep this from being a 
perfect equality.
The simplest directional correlation we can construct is ``up-down''. 
The neutron detectors were all placed below the mid-plane of EVA, 
so the neutrons were all detected in the ``downward'' direction. 
We can then compare this with the ``upward'' component of {\bf p$_f$}.

Figure \ref{PfxPn45h} is a scatter plot for (p,2p+n) events of 
p$_f^{up}$ vs. the neutron momentum 
p$_n$, where p$_f^{up}$ is the projection of {\bf p$_f$} on an axis normal 
to the faces of the neutron detectors. The vertical dashed line in 
Fig. \ref{PfxPn45h} is at k$_F$ = 0.22 GeV/c, the 
Fermi momentum for $^{12}$C \cite{Moniz}. 
Similar to \cite{YMar}, 
the experimental uncertainties in Fig. \ref{PfxPn45h} are derived from 
resolutions measured for p-p elastic scattering.
We note in Fig. \ref{PfxPn45h}, that there is a striking difference in the 
distribution of p$_f^{up}$ for p$_n$ $<$ k$_F$ and p$_n$ $\ge$ k$_F$. 
For p$_n$ $<$ k$_F$ the events are distributed nearly equally 
between those with p$_f^{up}$ $>$ 0 (``up'') and  p$_f^{up}$ $<$ 0 
(``down''). By contrast, for p$_n$ $\ge$ k$_F$, a large fraction of 
the events have p$_f^{up}$ $>$ 0. Thus {\bf p$_f$} is predominantly ``up'' 
when {\bf p$_n$} is (by definition) ``down'' for p$_n$ $\ge$ k$_F$. 
This is a strong 
confirmation of what was reported by Aclander et al. in \cite{acl99}.

We now construct the full directional correlation between 
{\bf p$_f$} and {\bf p$_n$} as
\begin{equation}
cos\gamma = {{\bf p_f} \cdot {\bf p_n} \over{\mid {\bf p_f} \mid \mid {\bf p_n} \mid} }
\end{equation}
where $\gamma$ is the angle between {\bf p$_f$} and {\bf p$_n$}.

\begin{figure} 
\vspace{-.8cm}
\centerline{\epsfxsize=.38\textwidth \epsfbox{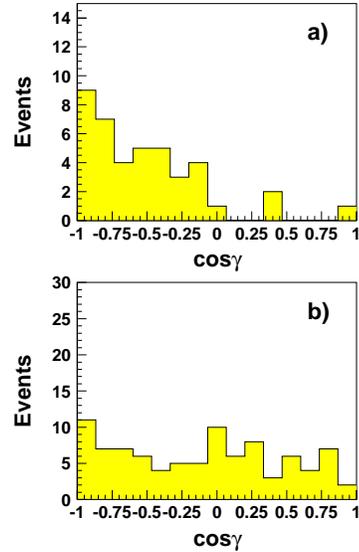}}
\vspace{-.7cm}
\caption{Plots of cos$\gamma$, where $\gamma$ is the angle between 
{\bf p$_n$} and {\bf p$_f$}, for $^{12}$C(p,2p+n) events. 
Panel (a) is for events with p$_n$ $>$ 0.22 GeV/c, 
and panel (b) is for events with p$_n$ $<$ 0.22 GeV/c; 0.22 GeV/c = k$_F$, 
the Fermi momentum for $^{12}$C. }
\label{cosgamma}
\vspace{.2cm}
\end{figure}

In Fig. \ref{cosgamma}a and \ref{cosgamma}b we plot cos$\gamma$ for p$_n$ $\ge$ k$_F$ = 0.22 GeV/c 
and p$_n$ $<$ k$_F$, respectively. As in Fig. \ref{PfxPn45h} there is 
a pronounced difference between the distributions for p$_n$ $\ge$ k$_F$ 
and p$_n$ $<$ k$_F$. Fig. \ref{cosgamma}a shows a strong ``back-to-back'' 
directional correlation with the distribution peaking at cos$\gamma$ = $-$1; 
only 3 events have cos$\gamma$ $>$ 0. By contrast, in Fig. \ref{cosgamma}b, 
the distribution is nearly uniform in cos$\gamma$, and the numbers of 
events for cos$\gamma$ $<$ 0 and cos$\gamma$ $>$ 0, are the same 
within statistics (50 and 40, respectively).
For Fig. \ref{cosgamma}a (p$_n$ $>$ k$_F$), the probability that 41 
uncorrelated events could be distributed 38 to 3 is vanishing 
small ($\sim$ 10$^{-8}$). To check the results in Fig. \ref{cosgamma}, 
we generated background spectra of events with 
two high p$_t$ tracks plus one soft track in EVA. The resulting 
spectra for both p$_n$ $>$ k$_F$ and p$_n$ $<$ k$_F$ were small and 
flat, indicating that the peak at cos$\gamma$ $= -$1 in Fig. \ref{cosgamma}a 
is not an instrumental artifact, and that backgrounds are small.

The effects of initial-state and final-state interactions (ISIs and FSIs) 
are expected to be larger in the transverse than in the longitudinal 
direction. Therefore, in extracting the relative and c.m. motion for 
correlated n-p pairs we focus on the longitudinal ({\em z}) components 
of {\bf p$_n$} and {\bf p$_f$}. 
For beams of high-energy protons, the natural variables 
for representing the motion of a nucleon in the nucleus are the 
light-cone variables {\bf p$_t$} and $\alpha$ = (E $-$ p$_z$)/m. Note 
that $\alpha$ $\approx$ 1 for p$_z$ = 0, therefore, we would expect 
$\alpha_f$ + $\alpha_n$ $\approx$ 2 for a correlated pair with 
{\bf p$_n$} $\approx$ $-${\bf p$_f$}. ($\alpha_f$ = (E $-$ p$_{fz}$)/m; 
 $\alpha_n$ = (E $-$ p$_{nz}$)/m)

\begin{figure}  
\vspace{-.7cm}
\centerline{\epsfxsize=0.32\textwidth \epsfbox{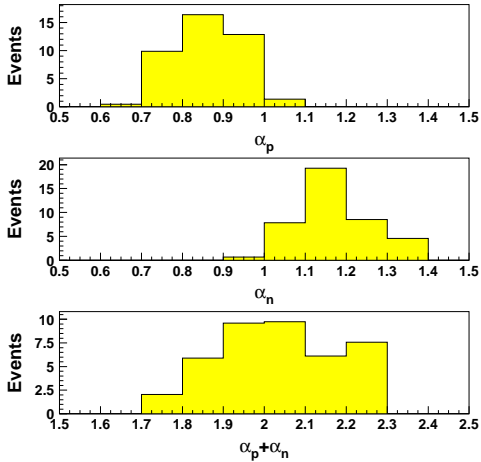}}
\vspace{-.7cm}
\caption{Plots of the light-cone variables $\alpha_p$, $\alpha_n$, and their 
sum $\alpha_p$ + $\alpha_n$ for events with p$_n$ $>$ k$_F$ = 0.22 GeV/c. Data 
are for $^{12}$C(p,2p+n) events. Each event 
has been ``{\em s}-weighted", as described in the text. }
\label{alphas_Gt}
\vspace{.2cm}
\end{figure}

As noted earlier, the p-p cross section, 
$d\sigma \over dt$,
near 90$^\circ$ c.m. is proportional to {\em s}$^{-10}$. We therefore need to 
correct for this ``reaction bias'' when looking at longitudinal variables.
The differential cross section as a function of the solid angle, 
$d\sigma \over d\Omega$, is related to $d\sigma \over dt$ as:
\begin{equation}
   {d\sigma \over dt} = { {4\pi}\over (s - 4m^2) }{d\sigma \over d\Omega}  
\end{equation}
Therefore, for large {\em s}, $d\sigma \over d\Omega$ is approximately 
proportional to {\em s}$^{-9}$, and this strong {\em s}-dependence 
enhances quasielastic reactions with low {\em s} for the p-p collisions. 
Protons in the nucleus with longitudinal momentum 
in the same direction as the beam are thus more likely 
to be knocked out. We therefore weighted each event by a correction 
factor equal to ({\em s/s$_0$})$^9$, to obtain the nuclear distributions 
without the reaction bias, where {\em s}$_0$ is the total c.m. energy for 
pp$\to $pp at each beam momentum, and {\em s} is calculated for each 
event from {\bf p$_f$} for that event.

In Figure \ref{alphas_Gt} we show plots of $\alpha_f$, $\alpha_n$ and
$\alpha_f$ + $\alpha_n$ (all with {\em s}-weighting) for events with p$_n$
$>$ k$_F$.  We note for Fig. \ref{alphas_Gt} that $\alpha_f$ (for the
proton) is generally $<$ 1 and $\alpha_n$ $>$ 1. Of course, our placement
of the neutron detectors primarily in the backward hemisphere forces
$\alpha_n$ to be largely $>$ 1. In Fig. \ref{alphas_Gt}, the spread of
$\alpha_n$ + $\alpha_f$ about 2 should be due to the c.m. motion of the
pair. Cioffi degli Atti {\it et al.} \cite{Cioffi} emphasized the
importance of the c.m.
motion of correlated pairs for explaining nucleon spectral functions
at large momenta and removal energies.

In the longitudinal direction:
\begin{equation}
   p_z^{cm} = p_{nz} + p_{fz}.  
\end{equation}
By approximating E$_p$ $\approx$ E$_n$ $\approx$ m, we obtain
\begin{equation}
 \alpha_p + \alpha_n = (1 - {p_{fz}\over m}) + (1 - {p_{nz}\over m})  
\end{equation}
which leads to
\begin{equation}
  p_z^{cm} = 2m(1 - {{\alpha_p + \alpha_n}\over 2}).  
\end{equation}
The longitudinal momentum of the 
particles in their c.m. frame can be extracted from the difference 
of the $\alpha$ variables. By again approximating 
E$_p$ $\approx$ E$_n$ $\approx$ m, we obtain
\begin{equation}
  \alpha_p - \alpha_n = (1 - {p_{fz}\over m}) - (1 - {p_{nz}\over m}) 
= ({{p_{nz} - p_{fz}}\over m}),  
\end{equation}
which leads to
\begin{equation}
  p_z^{rel} = m\vert \alpha_p - \alpha_n \vert . 
\end{equation}

\begin{figure}[] 
\vspace{-1.0cm}
\centerline{\epsfxsize=.35\textwidth \epsfbox{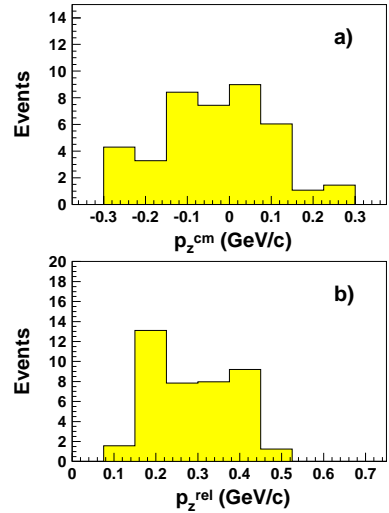}}
\vspace{-.7cm}
\caption{Plots of (a) p$_z^{cm}$ and (b) p$_z^{rel}$ for correlated 
n-p pairs in $^{12}$C, for $^{12}$C(p,2p+n) events. Each event has been ``{\em s}-weighted", as described in the text. }
\label{PcmzPrelz}
\vspace{.2cm}
\end{figure}

Figure \ref{PcmzPrelz}a and \ref{PcmzPrelz}b are plots of 
$p_z^{cm}$ and $p_z^{rel}$. 
For the data for p$_z^{cm}$ in Fig. \ref{PcmzPrelz}a, the centroid 
is $-$0.013$\pm$0.027 GeV/c.  
The spread in the distribution is $\sigma$ = 0.143$\pm$0.017 GeV/c. 
For the data for p$_z^{rel}$ in Fig. \ref{PcmzPrelz}b, the centroid 
is 0.289$\pm$0.017 GeV/c, 
and the width is $\sigma$ = 0.097$\pm$0.007 GeV/c.  


An interesting number which can be extracted from our data, is the 
fraction of $^{12}$C(p,2p) events which have correlated neutrons with 
{\bf p$_n$} $\approx$ $-${\bf p$_f$} when p$_n$, p$_f$ $\ge$ k$_F$. 
To extract this number, 
we need to correct the measured neutron flux for neutron detection efficiency 
and flux attenuation, and for solid-angle 
coverage. Our neutron detectors were placed almost entirely 
in the backward hemisphere, so we calculate the fraction of the 2$\pi$ 
solid angle for the backward hemisphere covered by our detectors.

What we then calculated was:
\begin{equation}
F = {{corrected \mbox{ \# {\it of (p,2p+n) events}}}\over {\mbox{\# {\it of (p,2p) events}}} }
= {A \over{B}}
\end{equation}
for the same data sample. B was obtained by 
applying cuts (1), (2), (3) 
and (5), to events with p$_f$ $\ge$ k$_F$ for our 5.9 GeV runs excluding 
the data reported 
in \cite{acl99}. The quantity B = 2205 then was all events satisfying 
the above cuts and data selection. The quantity A was obtained from 
the sample of all 
18 (p,2p+n) events in the sample B with p$_n$ $\ge$ k$_F$, 
where a correction for flux attenuation ({\em t}) \cite{hughes} and 
detection efficiency ($\epsilon$) \cite{madey} was 
applied event-by-event. The resulting quantity was then corrected for the 
solid-angle coverage to obtain A:
\begin{equation}
A = {{2\pi}\over{\Delta\Omega}} \sum_{i=1}^{18} {1 \over{\epsilon_i}} \cdot {1 \over{t_i}} = 1090.
\end{equation}
The average value of $\epsilon_i^{-1}${\em t}$_i^{-1}$ was 8.2$\pm$0.82 
and 2$\pi$/$\Delta\Omega$ = 7.42.  
We then obtain
\begin{equation}
F = {A \over{B}} = 0.49 \pm 0.13.
\end{equation}
The sample of 18 measured (p,2p+n) events is clearly small, and the 
uncertainty in F is determined largely by this sample size. The 
result is still compelling: roughly half of the measured quasielastic 
(p,2p) events with p$_f$ $>$ k$_F$ have a neutron emitted in the 
backward hemisphere with p$_n$ $>$ k$_F$. 
We note that this result is very similar to that reported 
by Malki, et al. \cite{Malki} for hard inclusive (p,2p+n) measurements.


In summary, for quasi-elastic (p,2p) events we reconstructed 
{\bf p$_f$}, the momentum 
that the struck proton had in $^{12}$C before the reaction. Then, 
for neutrons with momenta p$_n$ $>$ 0.220 GeV/c (which is the k$_F$, 
Fermi momentum for $^{12}$C), 
we found a strong directional correlation between {\bf p$_f$} and {\bf p$_n$}, 
namely {\bf p$_n$} $\approx$ $-${\bf p$_f$}. By contrast, for p$_n$ $<$ k$_F$ we 
found no correlation in directions. This was evident in the one-dimensional 
``up-down'' and longitudinal correlations and the full three-dimensional directional correlation. 
For the longitudinal direction, where ISI and FSI effects should be small, 
we extracted the distributions of c.m. and relative motion for n-p pairs.

We conclude, therefore, that neutrons emitted into the backward hemisphere 
with p$_n$ $>$ k$_F$ come from n-p SRC, since SRC is  
a natural mechanism to explain such momentum-correlated pairs. An analysis 
of the sample of events for 5.9 GeV/c beam momentum with p$_f$ $>$ k$_F$, 
indicates that 49 $\pm$ 13\% of these events have a correlated neutron with 
p$_n$ $>$ k$_F$. 
Because this measured fraction includes only n-p and not p-p SRC,  
the total correlated fraction must be even larger. Therefore we conclude that 
2N SRC must be a major source of high-momentum nucleons in nuclei.
We also measured the longitudinal
components of the c.m. momentum of the correlated pn 
pair and the relative momentum of the pn pair in
its c.m. system.
Further applications of 
the technique used in this work and in Ref. \cite{acl99} are planned \cite{prop} with the high-intensity c.w. 
electron beams at the Thomas Jefferson National Accelerator Facility.

\begin{acknowledgments}

We are pleased to acknowledge the assistance of 
the AGS staff in building and rebuilding the detector and supporting 
the experiment, particularly our liaison
engineer C. Pearson, and the continuing
support of  Drs. D. Lowenstein and P. Pile. 
This research was supported by the U.S. - Israel Binational Science
foundation, the Israel Science Foundation founded by the Israel Academy of
Sciences and Humanities, NSF grants No. PHY-9501114, PHY-9722519, PHY-0099387 
and the U.S. Department of Energy grant No. DEFG0290ER40553.

\end{acknowledgments}


\end{narrowtext}
\end{document}